\documentclass[prd,superscriptaddress,nofootinbib,amsmath,amssymb,aps,11pt]{revtex4}

\usepackage{bm}
\usepackage{amsfonts}
\usepackage{latexsym}
\usepackage[latin1]{inputenc}
\usepackage{graphicx}
\usepackage{amsmath}
\usepackage{palatino}
\usepackage{mathpazo}
\usepackage{booktabs}
\usepackage{dcolumn}

\def\jnl@style{\it}
\def\aaref@jnl#1{{\jnl@style#1}}

\def\aaref@jnl#1{{\jnl@style#1}}

\def\aj{\aaref@jnl{AJ}}                   
\def\apj{\aaref@jnl{ApJ}}                 
\def\apjl{\aaref@jnl{ApJ}}                
\def\apjs{\aaref@jnl{ApJS}}               
\def\apss{\aaref@jnl{Ap\&SS}}             
\def\aap{\aaref@jnl{A\&A}}                
\def\aapr{\aaref@jnl{A\&A~Rev.}}          
\def\aaps{\aaref@jnl{A\&AS}}              
\def\mnras{\aaref@jnl{Mon.~Not.~Roy.~Astron.~Soc.}}             
\def\prd{\aaref@jnl{Phys.~Rev.~D}}        
\def\prc{\aaref@jnl{Phys.~Rev.~C}}  
\def\prl{\aaref@jnl{Phys.~Rev.~Lett.}}    
\def\qjras{\aaref@jnl{QJRAS}}             
\def\skytel{\aaref@jnl{S\&T}}             
\def\ssr{\aaref@jnl{Space~Sci.~Rev.}}     
\def\zap{\aaref@jnl{ZAp}}                 
\def\nat{\aaref@jnl{Nature}}              
\def\aplett{\aaref@jnl{Astrophys.~Lett.}} 
\def\apspr{\aaref@jnl{Astrophys.~Space~Phys.~Res.}} 
\def\physrep{\aaref@jnl{Phys.~Rep.}}      
\def\physscr{\aaref@jnl{Phys.~Scr}}       
\def\commat{\aaref@jnl{Comm.~Math.~Phys.}}              
\def\science{\aaref@jnl{Science}}               
\def\cqg{\aaref@jnl{Classical Quant.~Grav.}}            
\def\jpcs{\aaref@jnl{JPCS}}                                     
\def\ijmpd{\aaref@jnl{Int.~J.~Mod.~Phys.~D}}                    
\def\grg{\aaref@jnl{Gen.~Relat.~Gravit.}}               
\def\rpp{\aaref@jnl{Rep.~Prog.~Phys.}}          
\def\npa{\aaref@jnl{Nucl.~Phys.~A}}        
\def\lrr{\aaref@jnl{Living Rev.~Rel.}}                   
\def\jcap{\aaref@jnl{J.~Cosmology Astropart.~Phys.}}    
\def\rmp{\aaref@jnl{Rev.~Mod.~Phys.}}   

\allowdisplaybreaks[1]

\begin{document}

	\title{Black hole scalarization induced by the spin -- 2+1 time evolution}
	
	\author{Daniela D. Doneva}
	\email{daniela.doneva@uni-tuebingen.de}
	\affiliation{Theoretical Astrophysics, Eberhard Karls University of T\"ubingen, T\"ubingen 72076, Germany}
	\affiliation{INRNE - Bulgarian Academy of Sciences, 1784  Sofia, Bulgaria}
	
	\author{Lucas G. Collodel}
	\email{lucas.gardai-collodel@uni-tuebingen.de}
	\affiliation{Theoretical Astrophysics, Eberhard Karls University of T\"ubingen, T\"ubingen 72076, Germany}

	\author{Christian J. Kr\"uger}
	\email{christian.krueger@tat.uni-tuebingen.de}
	\affiliation{Theoretical Astrophysics, Eberhard Karls University of T\"ubingen, T\"ubingen 72076, Germany}

	\author{Stoytcho S. Yazadjiev}
	\email{yazad@phys.uni-sofia.bg}
	\affiliation{Theoretical Astrophysics, Eberhard Karls University of T\"ubingen, T\"ubingen 72076, Germany}
	\affiliation{Department of Theoretical Physics, Faculty of Physics, Sofia University, Sofia 1164, Bulgaria}
	\affiliation{Institute of Mathematics and Informatics, 	Bulgarian Academy of Sciences, 	Acad. G. Bonchev St. 8, Sofia 1113, Bulgaria}


	\begin{abstract}
		The scalarization is a very interesting phenomena allowing to endow a compact object with scalar hair while leaving all the predictions in the weak field limit unaltered. In Gauss-Bonnet gravity the source of the scalar field can be the curvature of the space-time. It was recently shown that for a particular type of coupling function between the scalar field and the Gauss-Bonnet invariant, spin-induced black hole scalarization is possible. In the present paper we study this phenomenon by performing a 2+1 time evolution of  the relevant  linearized scalar field perturbation equation and examine the region where the Kerr black hole becomes unstable giving rise to new scalarized rotating black holes. This is more direct numerical approach to study the development of spin-induced scalarization and it can serve as an independent check of the previous results. 
	\end{abstract}
	
	\maketitle
	
	\section{Introduction}

	Alternative theories of gravity play a major role in the research performed by relativists and cosmologists around the globe, even while general relativity (GR) keeps thriving with its predictions as new ways of probing it are developed. Two main reasons support this practice. The first is our current understanding of the large scale structure of the universe based on observations, which requires the existence of a weakly interacting matter field known as dark matter, yet not experimentally found, as well as a negative pressure field which we call dark energy, capable of producing the late time accelerated expansion, and also a similar field named inflaton which would have caused the early inflationary period of our universe. The second reason is that GR is a classical non-renormalizable theory that is expected to break down at the Planck scale, and therefore cannot tell the whole story. In exploring different theories, one may not ignore the success achieved by GR so far, hence any deviations from it on classical scales are assumed to take place at the strong regime, i.e. in the vicinity of a compact object such as a black hole. Understanding what features to look for in such deviations and how they relate among different theories is the key to better categorize the whole myriad of generalizations found thus far in the literature.
	
	In GR and several classes of scalar-tensor theories the usual no-hair theorems apply, see i.e. \cite{PhysRevD.5.1239,PhysRevD.54.5059,Sotiriou_2015,PhysRevLett.108.081103,doi:10.1142/S0218271815420146}. The exceptions are given by dropping the theorem's assumption that the matter fields share the same isometries of the spacetime, which allows for black holes with \emph{synchronized hair} given by some complex spin-0 or spin-1 field endowed with a Noether charge that cannot be accessed asymptotically through the metric components \cite{PhysRevLett.112.221101,PhysRevD.99.064036,DELGADO2019436,DELGADO2016234,PhysRevD.92.084059,Herdeiro_2016,Santos:2020pmh,PhysRevD.100.064032,Kunz:2019sgn,Collodel:2020gyp}. Accordingly, the existence of hairy black holes greatly constrains the classes of theories which are serious contenders to expand GR's predictions.
	
	Extensively studied is a special subset of Hordenski theories known as Einstein-dilaton-Gauss-Bonnet (EdGB) theory, where the general relativistic action is extended to include quadratic terms which are curvature invariant -- the Gauss Bonnet term. Such a term alone is a topological one which does not contribute to the dynamics in four spacetime dimensions, but in the EdGB theory it is coupled to a scalar field through an exponential function $f(\varphi)=\alpha\exp(-\gamma\varphi)$ \cite{Guo:2008hf,Lee:2018zym,Gonzalez:2018aky,Cunha:2016wzk,Blazquez-Salcedo:2017txk}. The dynamics of the dilatonic field is prescribed by a Klein-Gordon (KG) equation (plus Maxwell's equations when further coupled to an electromagnetic field) that does not allow for trivial  (scalar hair free) solutions, due to the special form of the coupling. Therefore, dilatonic hair is featured in all solutions and GR black holes are not part of them.
	
	Recently, it has been realized that by modifying the coupling function it is possible to recover GR as a special set of the theory, in particular if $df(\varphi)/d\varphi=0$ for a certain constant $\varphi_0$, then $\varphi=\varphi_0$ everywhere is a possible solutions to the KG equations. The scalar field no longer represents the dilaton, and this class is generally referred to as Einstein-scalar-Gauss-Bonnet theory (EsGB). This constraint on the coupling function is interesting for it leads to a \emph{curvature induced spontaneous scalarization} as noted for the first time for static black holes \cite{PhysRevLett.120.131103,PhysRevLett.120.131104,PhysRevLett.120.131102}. In short, around compact objects the curvature becomes large enough to trigger the dynamics of the scalar field. Thereafter, different objects have been studied in this class of theories, such as spherically symmetric black holes with different couplings and their stability properties \cite{PhysRevD.97.084037,PhysRevD.99.044017,PhysRevD.99.064011,PhysRevD.99.064003,PhysRevD.99.104045,PhysRevD.99.104041,PhysRevD.98.104056,PhysRevD.101.024033, PhysRevD.98.084011,Blazquez-Salcedo:2020rhf,Blazquez-Salcedo:2020caw,Doneva:2020qww}, scalarized neutron stars \cite{PhysRevLett.120.131104,Doneva:2017duq}, wormholes with no exotic matter as the theories allow for the violation of the null energy condition (NEC) \cite{PhysRevD.101.024033} and even particle-like solutions \cite{KLEIHAUS2020135401,PhysRevD.102.024070}, which describe a scalar field divergent at the origin, but whose energy-momentum tensor is regular everywhere, and so is the spacetime. Rotating EsGB black holes, which are the topic of investigation of this article, have been reported in \cite{PhysRevLett.123.011101,Collodel_2020} for different coupling functions.
	
	In this paper we probe the stability of Kerr EsGB black holes by fully evolving in $2+1$ dimensions the modified Klein-Gordon  equation 
	describing the scalar perturbation of Kerr black holes within the EsGB gravity, and determine in which regions of the parameter space that defines the theory a tachyonic instability gives rise to hairy black holes.  We focus only on the case where the second order derivative of the coupling function has a negative sign that cannot lead to scalarization in the static case as in \cite{PhysRevLett.120.131103,PhysRevLett.120.131104}, but instead spin-induced scalarization is observed above a certain threshold of the black hole angular momentum. We note that a similar work has recently appeared in pre-print versions \cite{Dima:2020yac}, later supported by some analytical calculations \cite{Hod:2020jjy}, but it uses a different methodology where one decomposes the angular dependencies onto a basis set of spherical harmonics and evolves the resulting system in $1+1$ dimensions. Both works originally presented results that confronted our findings, which led to the development of this manuscript, but have now been updated on the pre-print server and the corrected versions are in agreement with our report.

	\section{Scalar field perturbations within Gauss-Bonnet gravity}	
	The action of Einstein-scalar-Gauss-Bonnet gravity in vacuum  is given by  
	
	\begin{eqnarray}
		S=&&\frac{1}{16\pi}\int d^4x \sqrt{-g} 
		\Big[R - 2\nabla_\mu \varphi \nabla^\mu \varphi 
		+ \lambda^2 f(\varphi){\cal R}^2_{GB} \Big] ,\label{eq:quadratic}
	\end{eqnarray}
	where, as usual, $R$ is the Ricci scalar with respect to the spacetime metric $g_{\mu\nu}$, $\varphi$ is the scalar field  with a coupling function  $f(\varphi)$, $\lambda$ is the Gauss-Bonnet coupling constant having  dimension of $length$ and ${\cal R}^2_{GB}$ is the Gauss-Bonnet invariant \footnote{Note that $\lambda$ is connected to the parameter $\eta$ used in \cite{Dima:2020yac} in the following way $\lambda^2=\frac{1}{4}\eta$}. The Gauss-Bonnet invariant is defined by ${\cal R}^2_{GB}=R^2 - 4 R_{\mu\nu} R^{\mu\nu} + R_{\mu\nu\alpha\beta}R^{\mu\nu\alpha\beta}$ where $R_{\mu\nu}$ is the Ricci tensor and $R_{\mu\nu\alpha\beta}$ is the Riemann tensor. The action yields the following field equations 
	\begin{eqnarray}\label{FE}
		&&R_{\mu\nu}- \frac{1}{2}R g_{\mu\nu} + \Gamma_{\mu\nu}= 2\nabla_\mu\varphi\nabla_\nu\varphi -  g_{\mu\nu} \nabla_\alpha\varphi \nabla^\alpha\varphi ,\\
		&&\nabla_\alpha\nabla^\alpha\varphi=  -  \frac{\lambda^2}{4} \frac{df(\varphi)}{d\varphi} {\cal R}^2_{GB},
	\end{eqnarray}
	where  $\nabla_{\mu}$ is the covariant derivative with respect to the spacetime metric $g_{\mu\nu}$ and  $\Gamma_{\mu\nu}$ is defined by 
	\begin{eqnarray}
		&&\Gamma_{\mu\nu}= - R(\nabla_\mu\Psi_{\nu} + \nabla_\nu\Psi_{\mu} ) - 4\nabla^\alpha\Psi_{\alpha}\left(R_{\mu\nu} - \frac{1}{2}R g_{\mu\nu}\right) + 
		4R_{\mu\alpha}\nabla^\alpha\Psi_{\nu} + 4R_{\nu\alpha}\nabla^\alpha\Psi_{\mu} \nonumber \\ 
		&& - 4 g_{\mu\nu} R^{\alpha\beta}\nabla_\alpha\Psi_{\beta} 
		+ \,  4 R^{\beta}_{\;\mu\alpha\nu}\nabla^\alpha\Psi_{\beta} 
	\end{eqnarray}  
	with 
	\begin{eqnarray}
		\Psi_{\mu}= \lambda^2 \frac{df(\varphi)}{d\varphi}\nabla_\mu\varphi .
	\end{eqnarray}
	
	In what follows we will consider  asymptotically flat spacetimes and  the case for which the cosmological value of the scalar field is zero, i.e. $\varphi_{0}=0$. Without loss of generality we can impose the following constraints on the 
	coupling function $f(\varphi)$: $f(0)=0$ and $\frac{d^2f}{d\varphi^2}(0)=\epsilon$ with $\epsilon=\pm 1$. Since the focus of the present paper
	paper is on spontaneous scalarization we impose one more condition on $f(\varphi)$, namely  $\frac{df}{d\varphi}(0)=0$ which is crucial for the spontaneous scalarization. Under this condition it is not difficult to see that the Kerr black hole solution is also 
	a solution to the EsGB gravity with a trivial scalar field  $\varphi=0$. As in the non-rotating case the question that arises is whether 
	the Kerr solution is stable within the framework of the bigger EsGB theory. Of course the stability in general depends on the parameters 
	of the Kerr solution, i.e. the mass $M$ and angular momentum  $a$ as well as on the Gauss-Bonnet coupling parameter $\lambda$.

	In order to study the stability of the Kerr black hole  we shall consider the perturbation of the Kerr solution within the framework of EsGB gravity. It is not difficult to see that when the coupling function satisfies the condition $\frac{df}{d\varphi}(0)=0$, the equations governing the perturbations of the metric $\delta g_{\mu\nu}$ are decoupled from the equation governing the perturbation $\delta \varphi$ of the scalar field. The equations for metric perturbations  are in fact the same as those in the pure Einstein gravity and therefore we shall focus only on the scalar field perturbations. The equation governing the 
	scalar perturbation is    	
	\begin{eqnarray}\label{PESF}
		\Box_{(0)} \delta\varphi + \frac{\epsilon}{4}\lambda^2  {\cal R}^2_{GB(0)} \delta\varphi=0, 
	\end{eqnarray} 
	where $\Box_{(0)}$ and ${\cal R}^2_{GB(0)}$ are the D'alambert operator and the Gauss-Bonnet invariant for 
	the Kerr geometry. 
	
	As discussed in the introduction, the Gauss-Bonnet invariant can act as a source of the scalar field and leads to black hole scalarization. For this purpose, the whole term with $ {\cal R}^2_{GB(0)}$ should be positive  in a certain region of the parameters space. This is clearly possible for $\epsilon=1$ that leads to  scalarization of both static and rotating black holes recently discovered in \cite{PhysRevLett.120.131103,PhysRevLett.120.131104}. If $\epsilon=-1$, though, the term can remain positive if $ {\cal R}^2_{GB(0)}$ is greater than zero in a certain region outside the black hole. This is possible only for the rotating Kerr solution as we will demonstrate below. Thus, in the present paper we will focus only on the latter case, i.e. $\epsilon=-1$.
	
	In the standard Boyer-Lindquist coordinates,  the Kerr metric writes   
	\begin{eqnarray}\label{KerrM}
		ds^2= - \frac{\Delta -a^2\sin^2\theta}{\Sigma} dt^2 - 2a \sin^2\theta \frac{r^2 + a^2 - \Delta}{\Sigma} dt d\phi  
		+ \frac{(r^2 + a^2)^2 - \Delta a^2 \sin^2\theta}{\Sigma} \sin^2\theta d\phi^2 + \frac{\Sigma}{\Delta} dr^2 + \Sigma d\theta^2
	\end{eqnarray} 
	where $\Delta=r^2 - 2Mr + a^2$ and $\Sigma=r^2 + a^2 \cos^2\theta$. The Gauss-Bonnet invariant for the Kerr solution is explicitly given by 
	\begin{eqnarray}
		{\cal R}^2_{GB(0)}= \frac{48 M^2}{\Sigma^6}(r^2- a^2\cos^2\theta)(r^4 - 14 a^2 r^2 \cos^2\theta + a^4\cos^4\theta).
	\end{eqnarray} 
	
	Before writing the explicit form of the perturbation equation (\ref{PESF}) we introduce a new azimuthal coordinate $\phi_*$ defined by 
	\begin{eqnarray}
		d\phi_*=d\phi + \frac{a}{\Delta} dr.  
	\end{eqnarray} 
	Working with $\phi_*$ allows us to get rid of some unphysical pathologies near the horizon. It is also convenient to work with the tortoise coordinate $x$ given by 
	
	\begin{eqnarray}
		dx= \frac{r^2+a^2}{\Delta} dr.  
	\end{eqnarray} 
	
	In the coordinates $(t,x,\theta,\phi_*)$ the perturbation equation (\ref{PESF}) takes the following explicit form
	
	\begin{eqnarray} \label{eq:PertEq}
		&&-\left[(r^2 + a^2)^2 - \Delta a^2 \sin^2\theta\right] \partial^2_t \delta\varphi + (r^2 + a^2)^2 \partial^2_x \delta\varphi + 2r \Delta \partial_x\delta\varphi - 4Ma r\partial_t\partial_{\phi_*}\delta\varphi \nonumber \\ 
		&&+  2a(r^2 + a^2)\partial_x\partial_{\phi_*}\delta\varphi  + \Delta\left[\frac{1}{\sin\theta} \partial_\theta(\sin\theta\partial_\theta\delta\varphi) +   \frac{1}{\sin^2\theta}\partial^2_{\phi_*}\delta\varphi \right] \\ 
		&& = \lambda^2 \frac{12 M^2\Delta}{\Sigma^5}(r^2- a^2\cos^2\theta)(r^4 - 14 a^2 r^2 \cos^2\theta + a^4\cos^4\theta)\delta\varphi. \nonumber
	\end{eqnarray} 
	For large enough $a$, there exists a region of the parameter space 	where the term in the right-hand side is negative that can potentially 	lead to scalarization. While at the horizon the whole term is always 	zero, in its vicinity several negative local minima appear as one approaches the poles, $\theta=0$ and $\theta=\pi$. The threshold for 	the existence of negative part of the potential for these two points is $a=0.5$ and the negative minimum gets deeper by increasing $a$.  The appearance of such negative minima close to the poles is required but not sufficient condition to have unstable modes of the Kerr black hole and thus scalarization. Based on the analysis of the scalarization in the nonrotating case \cite{PhysRevLett.120.131103,PhysRevLett.120.131104}, though, one can expect that scalarization will develop quickly with the increase of $a$ and the consequent increase of the negative part of the term in the right-hand side. Indeed, the numerical simulations confirm this	observation as we will see in the next section.
	  
	\section{Numerical method}
	\subsection{Numerical approach}
	Clearly, eq. \eqref{eq:PertEq} is an analog of the Klein-Gordon equation  with variable  squared mass proportional to the Gauss-Bonnet term. Even though this new term is responsible for very important phenomenology, that is the destabilization of the Kerr solution and the appearance of scalarized hairy black holes, from a pure numerical point of view it does not bring additional serious challenges. Therefore, we can use the same approaches as for the numerical solution of the standard wave and Klein-Gordon equation. As mentioned in the Introduction, the approach we will follow is to directly evolve the perturbation eq. \eqref{eq:PertEq} in time, which, after separating out the azimuthal dependence, has only two spatial dimensions. We can perform this separation since the background is axisymmetric and we are considering only small perturbations on this background. Thus, we can assume the following form of the scalar field perturbation 
    \begin{equation}\label{eq:phi_ansatz}
	    \delta \varphi (t,x,\theta,\phi_*) = \delta \varphi (t,x,\theta) e^{im\phi_*},
    \end{equation}
    where $m$ is an integer -- the well-known azimuthal mode number. After substituting the above equation into eq. \eqref{eq:PertEq}, the $\phi_*$-derivative will be replaced by a simple multiplication with $im$, i.e., $\partial_{\phi_*} \rightarrow im$. This allows us to select a particular value for $m$ for any simulation.
	
	A step further was taken in \cite{Dima:2020yac} (see also \cite{Dima:2020rzg,Racz:2011qu}) where the angular dependence of the perturbation function $\delta \varphi$ was extended in series of the spherical harmonics that makes it possible to separate modes with different $l$ number. Since we have rotating solutions, it is not possible to rigorously attribute an $l$ number to a particular mode since the different $l$ modes couple to each other. Nevertheless, the experience shows that even for rapid rotation it is possible to define with a good accuracy $l$-led modes that in the limit $a=0$ reduce to the corresponding  nonrotating modes\footnote{In the nonrotating case the modes with different $l$ are decoupled and thus it is possible to rigorously attribute an $l$ number to every mode.} with the same $l$. This approach has the advantage that it is easier to separate the behavior for different $l$ and even to study the coupling between the different modes due to rotation. When solving eq. \eqref{eq:PertEq}, though, one is inevitably limited to a finite number of spherical harmonics in the expansion of $\delta \varphi$. Lower $l$ modes are normally those that are prone to instabilities so such a treatment of the problem is justified
	
	For this paper, we have opted for the time evolution approach in order to investigate the threshold for destabilization of the Kerr solutions. When evolving the $2+1$ version of the modified Teukolsky equation, it is possible, after properly choosing the initial data, to separate well the different  $l$ modes only in the nonrotating case where they decouple. As soon as the rotation is turned on, the mode coupling causes excitation of modes with different $l$ and it is expected that the $l=|m|$ mode will have the dominant contribution at late times if we limit ourselves to stable modes (see e.g. \cite{Gao:2018acg}). As far as unstable modes are concerned, the mode that is ``most unstable''  (meaning it has the shortest growth time)  for a fixed $m$ and for all $l$, will eventually dominate the signal. 
	
	We have to impose boundary conditions when evolving Eq. (11) in time. Both at the outer edge of the grid, being preferably as ``far away '' as possible, and at the inner edge, located very close to the event horizon, we impose usual outgoing wave boundary conditions (also known as Sommerfeld boundary condition) in order to remove energy that reaches those boundaries from the grid. Truncation errors will inevitably lead to undesired reflections from the outer boundary. One possibility to prevent the reflected wave from spoiling the extracted signal is to push the outer boundary of the numerical grid sufficiently far away, such that the spurious reflection returns to the location at which the signal is extracted only after a time such that the unspoiled signal has a long enough duration for a satisfactorily precise analysis.
	
	\subsection{Code implementation}
	The direct solution of the $2+1$ Teukolsky equation for different spins  was performed in  \cite{Krivan:1996da,Krivan:1997hc,Zenginoglu:2010cq,Harms:2013ib,Harms:2014dqa}. In addition, the modified KG equation in Chern-Simons gravity was considered in \cite{Gao:2018acg}, where the destabilization of the Kerr black hole was shown to appear for a certain range of parameters.  There are various sophisticated numerical techniques and analytical transformations that can be applied in order to cure different numerical problems like instabilities caused by the appearance of first order derivative in time or the reflection at the outer boundary. It turns out, though, that most of these problems are missing in our case or can be easily avoided. For example, in the modified Klein-Gordon equation \eqref{eq:PertEq} no first order time derivative is present unlike the Teukolsky equation with a spin different from zero. In addition, the time scale of the development of the instability is short enough and it is computationally affordable to push away the outer boundary in radial direction to large values so that the undesired reflected wave from infinity does not influence the observed signal, while still preserving a good accuracy. For that purpose we have chosen to implement a more straightforward numerical scheme that has shown to be robust enough for the problem of interest. 
	
	We are evolving in time equation \eqref{eq:PertEq} assuming the ansatz for the scalar field \eqref{eq:phi_ansatz}.  The numerical scheme is similar to the one considered in \cite{,Kruger:2019zuz,Kruger:2020ykw} for the evolution of spacetime perturbations around neutron stars.  For the calculation of spatial derivative we use a finite difference scheme with second order accuracy, while the integration in time is performed with a 3rd order Runge-Kutta solver. The calculations are performed on a grid that is uniform in the tortoise coordinate $x$ and the computational domain $[r_+ + \varepsilon_r, r_\infty]$, where $\varepsilon_r$ is a small shift from the outer horizon radius and $ r_\infty$ is our numerical infinity, is mapped to the tortoise coordinate domain $[x_{-\infty}, x_{+\infty}]$. The outgoing (ingoing) boundary conditions at $x_{\infty}$ ($x_{-\infty}$) is independent of the $\theta$ coordinate and reduces to the standard one $\partial_t \delta \varphi + \partial_x \delta \varphi = 0$ ($\partial_t \delta \varphi - \partial_x \delta \varphi = 0$) respectively. The numerical implementation of these boundary conditions follows \cite{RuoffPhD}.
	
	For the majority of the simulations we use the following parameters. The resolution of the grid is $500 \times 60$ in $x$ and $\theta$ directions. The numerical infinity is set to be at 50 horizon radii outside the black hole and the observer is located at roughly 30 horizon radii. Our results show that this leads to a maximum error or $0.5\%$ in the determination of the growth time of the unstable models while the threshold values of the parameters where instability develops has an accuracy over $99.9\%$. The reason we have chosen such moderate accuracy and value of $x_{\infty}$  is that some of the plots require to run thousands of simulation and the chosen parameters lead to a good balance between speed and accuracy.
	
	In the radial direction, the initial data has the form of a Gaussian pulse located at roughly $x_{\rm Gauss}=12$ and with size $\sigma=1$. The $\theta$ dependence of the result is taken to be a spherical harmonic with certain $l$. Even though the perturbation equation is clearly independent of $l$, it turns out that such initial data excite with a good accuracy predominantly modes with specific $m$ and $l$, especially in the nonrotating case where we lack the mode coupling.
	
	The oscillation frequencies are extracted after performing a Fourier transform of the signal. Inevitably, in most case this is connected with somewhat larger errors because of the limited number of observed oscillations before the asymptotic tail appears. This number of oscillation can vary from 2-3 in the case of $l=0$ initial data (i.e. $l=0$-led modes) and increases very rapidly with the increase of $l$ thus leading to better accuracy of the extracted frequency. The damping times are extracted by matching the peaks of the oscillation modes with an exponential function that is also prone to larger errors when a smaller number of peaks is present in the signal.
	
	It is clear that eq. \eqref{eq:PertEq} can be scaled with the black hole mass and thus the distance, the frequencies and the parameter $\lambda$ are measured in units of $M$ (that is numerically equivalent to just setting $M=1$ in this equation). Our calculations are performed in such dimensionless units.
	
	\section{Results}
	\subsection{Reliability of the numerical code}
	We have performed various tests in order to prove the correctness and to study the accuracy of the code we have developed. The first and most natural one is to calculate the quasinormal mode (QNM) frequencies of Schwarzschild and Kerr black holes. QNMs with specific $l$ and $m$ can be excited by simply choosing the $\theta$ dependence to have the form of the corresponding spherical harmonic. Our results show, that the obtained oscillation frequencies and damping times of Schwazschild and Kerr black holes are in agreement with the ones available in the literature (see e.g. \cite{Berti:2009kk,Berti:2005ys}) with an error typically up to a few percent that decreases for higher $l$.
	
	As a next step we have tried to confirm the results for the instabilities in  Chern-Simons gravity \cite{Gao:2018acg}. We have succeeded to reproduce with a very good accuracy the instability line, i.e. the threshold combination of $a$ and the Chern-Simons parameter $\eta_{CS}$ where the modes start to grow exponentially with time.
	
	Last but not least, we have tested the behavior of the code against different resolutions and dependence on the auxiliary parameters, such as the position of the numerical infinity and the width/location of the Gaussian pulse. We have verified numerically that the relevant quantities, such as the frequencies and the damping/growth time of the modes, indeed saturate to specific values with the increase of the resolution. The location of the numerical infinity does not have a direct relation to the accuracy of the calculated waveforms, but instead it spoils the signal after a fixed amount of time when the reflected signal from infinity reaches the point of signal extraction. Therefore, low values of $x_{+\infty}$ will effectively lead to a significant reduction of the number of clear oscillation cycles observed. This is not the case, though, for unstable modes, where the instability typically develops rapidly enough and thus $x_{+\infty}$  can be safely pushed towards much lower values than the ones required for proper waveform extraction of the stable modes.

	The observed signal is practically independent of the initial perturbation if we keep the width of the pulse in a reasonable range. For example a too narrow pulse will require also a large number of grid points in order to resolve it. We have chosen to work with a Gaussian pulse with $\sigma=1$ located at $x_{\rm Gauss}=12$. We have verified that our results are indeed independent of $\sigma$ and $x_{\rm Gauss}$ if we keep them in a reasonable range.
	
	\subsection{Instabilities of the scalar field perturbations}
	Due to the harmonic dependence the scalar field has on the axial coordinate, only the azimuthal mode number $m$ enters the perturbation equation explicitly. Thus for unstable models the exponential growth will be dominated by the fastest growing mode for a fixed $m$ independent of the initial perturbation. We have checked this explicitly and indeed the mode growth time is independent of the number $l$ of the initial perturbations. 
	
\begin{figure}[htp]
		\centering
		\includegraphics[width=.45\textwidth]{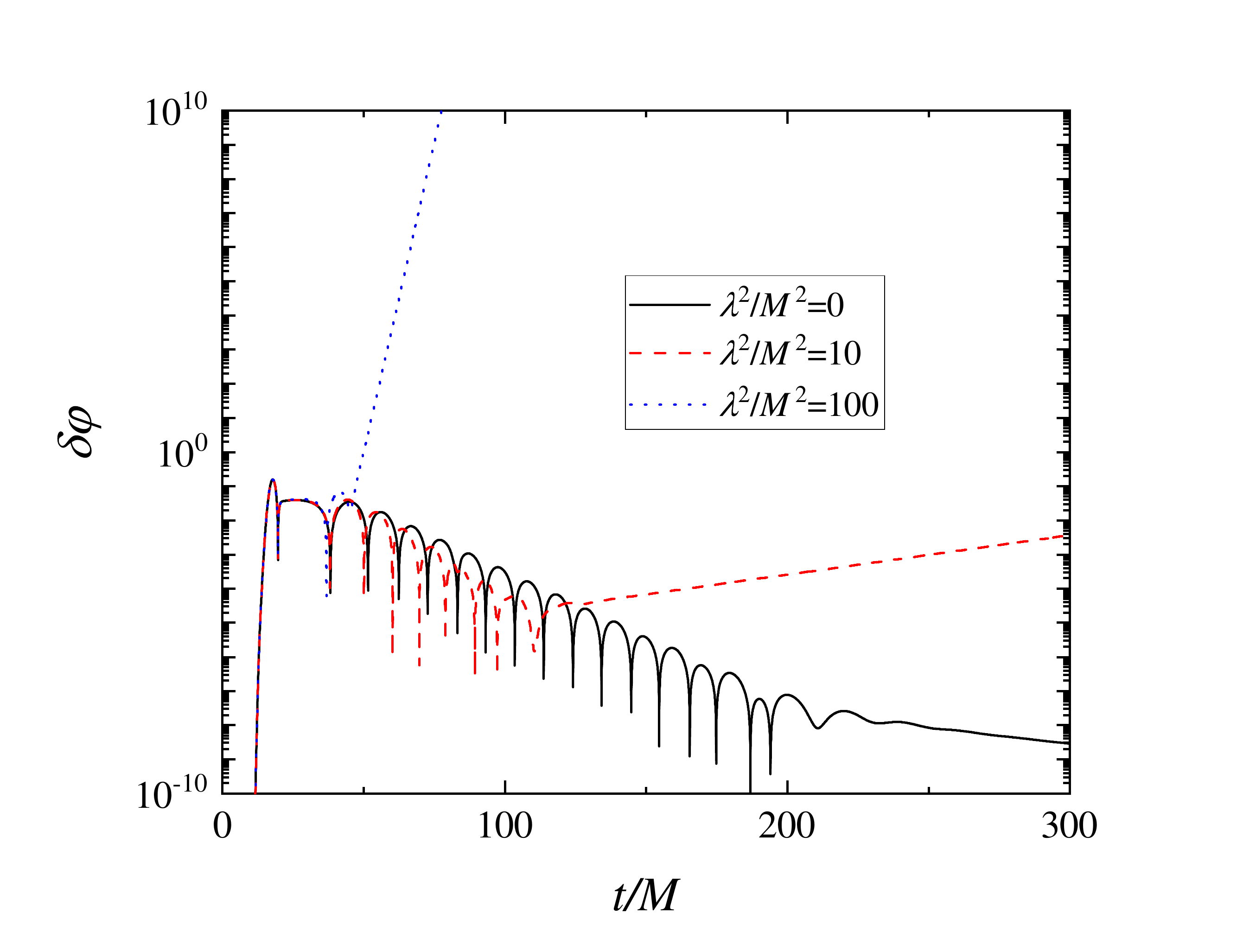}
		\includegraphics[width=.45\textwidth]{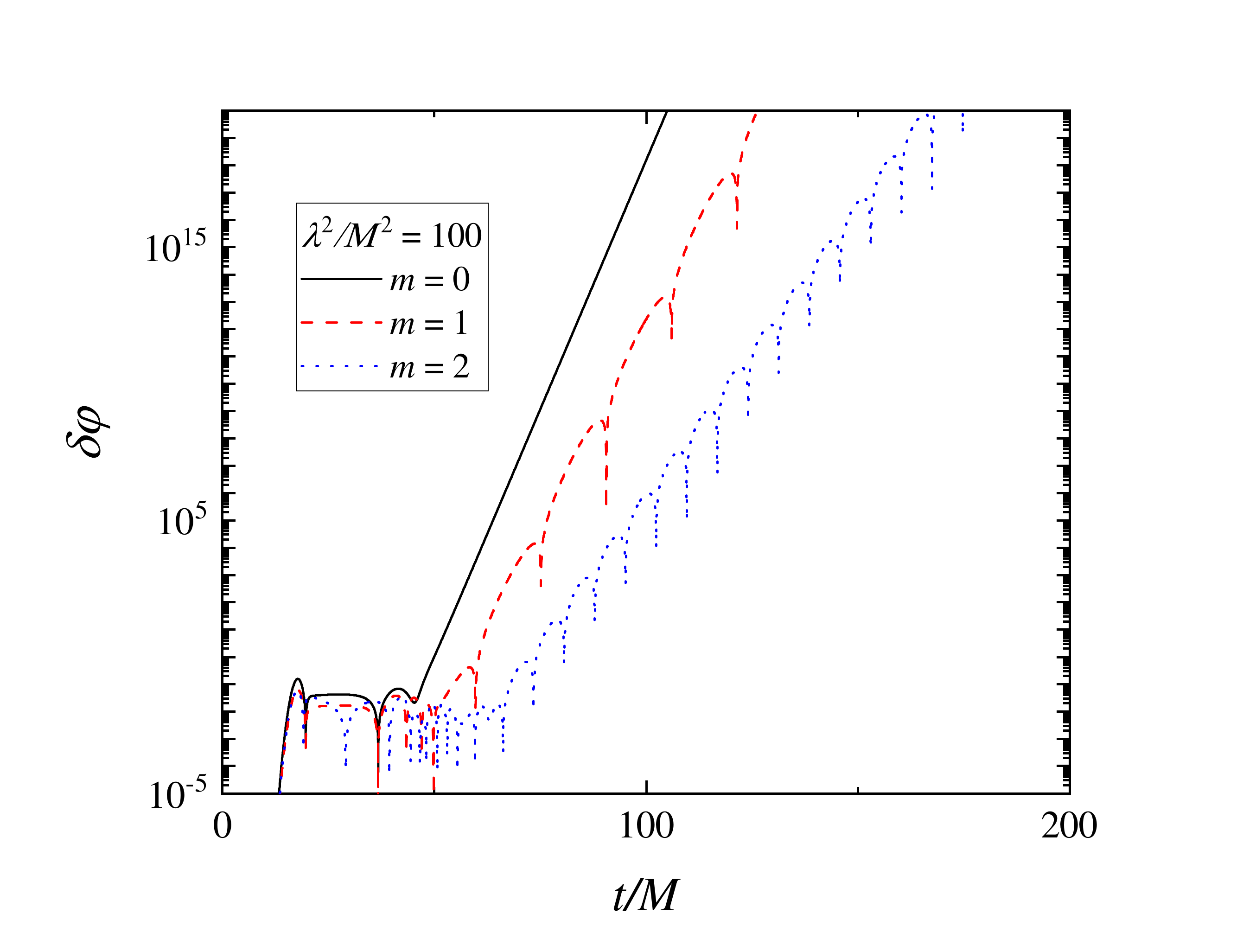}
		\caption{The signal $\delta \varphi$ for $m=0$ and different $\lambda$ (left figure) and $\lambda^2/M=100$ and several $m$ (right figure). The dimensionless angular moment is chosen to be $a/M=0.8$. The initial perturbation for $m=0$ and $m=1$ has $l=1$, while $l=2$ for $m=2$.  }
		\label{fig:exp_growth}
	\end{figure}

	In Fig. \ref{fig:exp_growth} (left panel) we show the behavior of the $m=0$ scalar field perturbation observed at $x=30$ for $a=0.8$ and several different value of $\lambda$ starting from pure GR with $\lambda=0$. The simulations are obtained using relatively large numerical infinity  compared to the rest of the simulations while keeping the step in $x$ direction constant,   in order to have a better visualization of the stable mode. More specifically, we use $2500\times 60$ points in $r$ and $\theta$ directions respectively, the numerical infinity is at $x_{+\infty}=250$. The initial data for $\delta \varphi$ has different values of $l$ for the different $m$, but as previously stated, this is relevant only for the stable modes or for the few oscillations observed before the start of the exponential growth of the mode for higher $\lambda$. For these calculations $m=0$ and therefore it is not possible to have supperradiant instability and the exponential growth is solely due to the instabilities connected to the scalarization. As one can see from the figure, the increase of the parameter $\lambda$ leads to a smaller growth time $\tau$ defined as $\delta \varphi \sim \exp{(t/\tau)}$, i.e. as expected higher $\lambda$ models develop the instability much more rapidly.
	
	The qualitative change of the signal for different $m$ is depicted in the right panel of Fig. \ref{fig:exp_growth}. As one can see, during exponential growth for $m=0$ no oscillations are present after the onset of instability. This is due to the fact that $m=0$ is a special case for which the real and the imaginary part of eq. \eqref{eq:PertEq} decouple\footnote{Real and imaginary $\delta \varphi$ can be introduced because of the substitution \eqref{eq:phi_ansatz} that results in a double number of equations for the real $\delta \varphi_R$ and the imaginary $\delta \varphi_I$.}. This is not the case, though, for $m>0$ where the coupling between real and imaginary  $\delta \varphi$ leads to oscillations even in the exponentially growing part of the signal (similar considerations but within the Chern-Simons gravity can be found in \cite{Gao:2018acg}). From this figure it is also evident that the growth time increases with the increase of $m$. This gives us the confidence to believe that the $m=0$ case leads to the shortest growth times and it is the most relevant one for the instabilities.

\begin{figure}[htp]
		\centering
		\includegraphics[width=.8\textwidth]{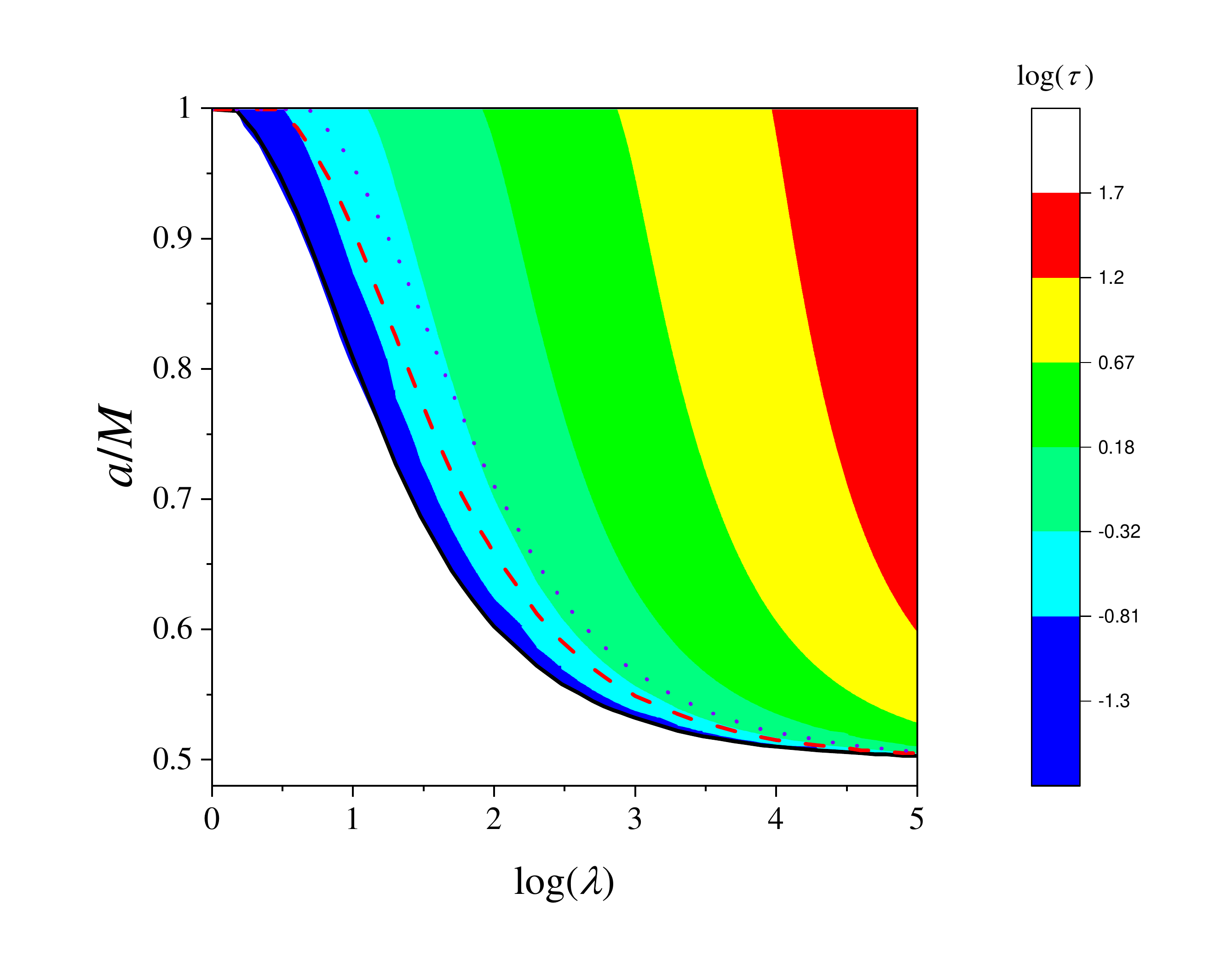}
		\caption{A contour plot of the unstable mode growth time for $m=0$. In addition , the instability lines, i.e. the threshold dependence $a(\lambda)$ where instability is lost, is plotted with solid black line for $m=0$, with a dashed red line for $m=1$ and with a dotted purple line for $m=2$.}
		\label{fig:contour}
	\end{figure}

	The contour plot of the growth times for $m=0$, that is supposed to be the most unstable case, is shown in Fig. \ref{fig:contour}. The threshold lines for the appearance of unstable Kerr solutions is plotted for $m=0,1,2$ and it is evident that for all fixed $a$, instabilities develop for larger critical values of $\lambda$ as $m$ is increased. In addition, one can notice that with the increase of $\lambda$, the threshold $a$ for the appearance of instabilities saturates for every $m$. This is expected since we have shown in the previous section that negative part of the right-hand side of the perturbation equation is possible only for $a>0.5$. In  \cite{Dima:2020yac} the threshold value for the development of instabilities  $a_{\rm crit}/M=0.505$ was found, for all $\lambda$ while in \cite{Hod:2020jjy} analytical considerations led to the conclusion that $a_{\rm crit}/M=0.5$ in the $l \rightarrow \infty$ limit.  The small discrepancy between the two results might be in the fact that the expansion of $\delta \varphi$ in spherical harmonics was truncated at a fixed $l$ and the it is practically impossible to clearly check the limit $l \rightarrow \infty$. As explained, our code does not suffer from such limitations, because only $m$ is an input parameter. Therefore, if an instability is present for any $l$ it will eventually appear and cause the amplitude to grow exponentially. Numerical instabilities appear, though, in our code for very large $\lambda^2/M^2$ of the order of $10^4-10^5$ that require an increase of the resolution. In addition, the closer we get to the critical value, the larger $\tau$ becomes and, therefore, we will need a very long evolution in order to check for instabilities. That is why the calculation for large value of $\lambda$ are very computationally demanding. 
	
\begin{figure}[htp]
		\centering
		\includegraphics[width=.8\textwidth]{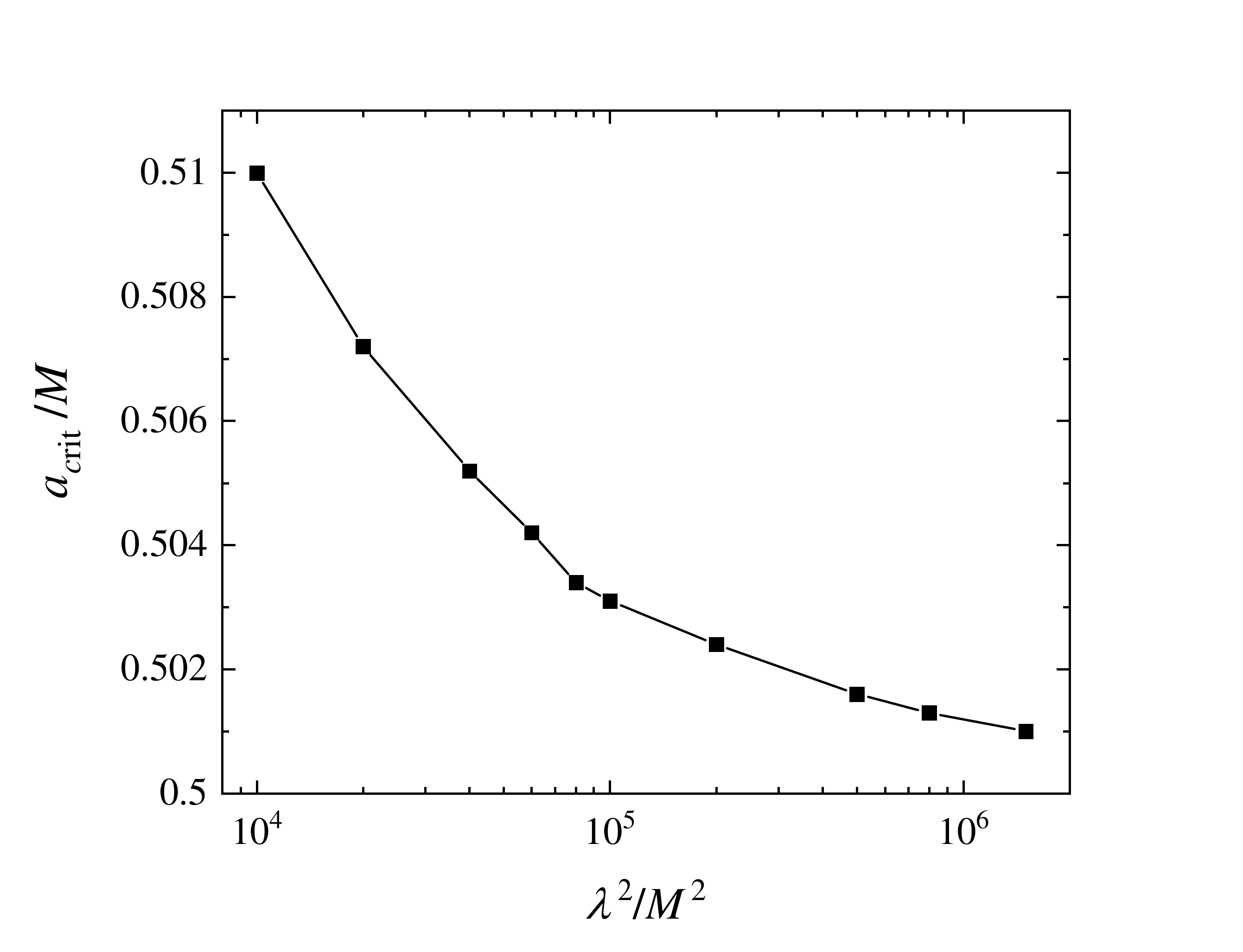}
		\caption{The critical value of $a_{\rm crit}$ where instability appears as a function of $\lambda^2/M^2$, for $m=0$. Only the region for very large $\lambda^2/M^2$ above $10^4$ is shown that corresponding to the bottom-left corner of Fig. \ref{fig:contour}.}
		\label{fig:acrit_zoom}
	\end{figure}

	In our calculations we reached up to $\lambda^2/M^2 \sim 10^6$ and in this case the critical value fall below $a_{\rm crit}/M \sim0.501$. The dependence of $a_{\rm crit}$ as a function of $\lambda^2$ (for very large $\lambda$) is given in Fig. \ref{fig:acrit_zoom} and it is evident that with increase of the coupling parameter, $a_{\rm crit}$ decreases but also starts to saturate. For the last points in Fig. \ref{fig:acrit_zoom} we had to increase the resolution by a factor of 5 in both $r$ and $\theta$ direction, that slows down the calculations a lot. From all this we can conclude that, despite the severe numerical instabilities appearing for large $\lambda$, the 2+1 calculations seem to be in agreement the $a_{\rm crit}/M=0.5$ limit derived  in \cite{Hod:2020jjy}. Another independent check of $a_{\rm crit}$ will be the construction of the actual equilibrium rotating scalarized solutions.
	
\section{Conclusions}
	In the present paper we have considered the 2+1 time evolution of the linearized scalar field perturbation equation in Gauss-Bonnet gravity. We are working with a subclass of theories having a negative second order derivative of the scalar field coupling function that can lead to spin-induced black hole scalarization as shown in \cite{Dima:2020yac,Hod:2020jjy}. In this case, for $a>0.5$, the Gauss-Bonnet term in the scalar field perturbation equation changes sign that leads to the appearance of negative parts in the effective two-dimensional potential close to the black hole horizon and rotational axes. In comparison with previous numerical studies we are performing the time evolution in two spacial dimensions without assuming any pre-defined form for the $\theta$ dependence. This approach is more direct and advantageous especially for the unstable modes since the final results for the (not) presence of instability is completely independent of the $l$ number and depends only on the azimuthal mode number $m$. Thus, it serves as in independent check of \cite{Dima:2020yac,Hod:2020jjy}.
	
	The developed numerical code is tested against different particular cases, such as the QNMs of Kerr black holes and the similar development of instabilities in Chern-Simons gravity. The code showed a good behavior when studying different  resolutions, positions of the numerical infinity and forms of the Gaussian pulse initial data. A general feature of the signal for unstable modes of black hole with large spin parameter $a$, is that after a few damped oscillation the amplitude of the scalar field perturbation starts to grow exponentially. For the marginally unstable modes, the time needed for the development of this instability might increase significantly and in addition the calculations require high resolution, that is another challenge to the 2+1 numerical evolution. 
	
	We have studied the instability region on a $a(\lambda)$ diagram, where  $\lambda$ is the Gauss-Bonnet coupling constant, for $m=0,1,2$. The results show that, as expected, the $m=0$ perturbations are the most relevant for the instability, i.e. they lead to the largest instability window. Naturally, the growth time of the unstable modes is shorter for larger $a$ and for every $\lambda$ there exist a critical angular momentum $a_{\rm crit}$ above which the Kerr black hole is unstable within the Gauss-Bonnet theories. A debate between \cite{Dima:2020yac,Hod:2020jjy} occurred about the exact value of  $a_{\rm crit}$ as $\lambda\rightarrow \infty$. Despite the severe numerical difficulties we had in the calculation for very large $\lambda$, our studied seems to confirm the analytical findings in \cite{Hod:2020jjy} about the minimum possible $a_{\rm crit}/M=0.5$. A very important independent check of this threshold angular momentum will be also the calculations of the stationary scalarized black hole solutions.
	  
	\section*{Acknowledgements}
	DD and LC acknowledges financial support via an Emmy Noether Research Group funded by the German Research Foundation (DFG) under grant
	no. DO 1771/1-1. DD is indebted to the Baden-Wurttemberg Stiftung for the financial support of this research project by the Eliteprogramme for Postdocs.  SY would like to thank the University of Tuebingen for the financial support.  
	The partial support by the Bulgarian NSF Grant KP-06-H28/7 and the  Networking support by the COST Actions  CA16104 and CA16214 are also gratefully acknowledged. C.K. acknowledges support from the DFG research grant 	413873357.
	
	
    \bibliographystyle{ieeetr}
	\bibliography{references,biblio_intro}

\begin{thebibliography}{10}

\bibitem{PhysRevD.5.1239}
J.~D. Bekenstein, ``Nonexistence of baryon number for static black holes,''
  {\em Phys. Rev. D}, vol.~5, pp.~1239--1246, Mar 1972.

\bibitem{PhysRevD.54.5059}
A.~E. Mayo and J.~D. Bekenstein, ``No hair for spherical black holes: Charged
  and nonminimally coupled scalar field with self-interaction,'' {\em Phys.
  Rev. D}, vol.~54, pp.~5059--5069, Oct 1996.

\bibitem{Sotiriou_2015}
T.~P. Sotiriou, ``Black holes and scalar fields,'' {\em Classical and Quantum
  Gravity}, vol.~32, p.~214002, oct 2015.

\bibitem{PhysRevLett.108.081103}
T.~P. Sotiriou and V.~Faraoni, ``Black holes in scalar-tensor gravity,'' {\em
  Phys. Rev. Lett.}, vol.~108, p.~081103, Feb 2012.

\bibitem{doi:10.1142/S0218271815420146}
C.~A.~R. Herdeiro and E.~Radu, ``Asymptotically flat black holes with scalar
  hair: A review,'' {\em International Journal of Modern Physics D}, vol.~24,
  no.~09, p.~1542014, 2015.

\bibitem{PhysRevLett.112.221101}
C.~A.~R. Herdeiro and E.~Radu, ``Kerr black holes with scalar hair,'' {\em
  Phys. Rev. Lett.}, vol.~112, p.~221101, Jun 2014.

\bibitem{PhysRevD.99.064036}
Y.-Q. Wang, Y.-X. Liu, and S.-W. Wei, ``Excited kerr black holes with scalar
  hair,'' {\em Phys. Rev. D}, vol.~99, p.~064036, Mar 2019.

\bibitem{DELGADO2019436}
J.~F. Delgado, C.~A. Herdeiro, and E.~Radu, ``Kerr black holes with
  synchronised scalar hair and higher azimuthal harmonic index,'' {\em Physics
  Letters B}, vol.~792, pp.~436 -- 444, 2019.

\bibitem{DELGADO2016234}
J.~F. Delgado, C.~A. Herdeiro, E.~Radu, and H.~Runarsson, ``Kerr-newman black
  holes with scalar hair,'' {\em Physics Letters B}, vol.~761, pp.~234 -- 241,
  2016.

\bibitem{PhysRevD.92.084059}
C.~A.~R. Herdeiro, E.~Radu, and H.~R\'unarsson, ``Kerr black holes with
  self-interacting scalar hair: Hairier but not heavier,'' {\em Phys. Rev. D},
  vol.~92, p.~084059, Oct 2015.

\bibitem{Herdeiro_2016}
C.~Herdeiro, E.~Radu, and H.~R{\'{u}}narsson, ``Kerr black holes with proca
  hair,'' {\em Classical and Quantum Gravity}, vol.~33, p.~154001, jun 2016.

\bibitem{Santos:2020pmh}
N.~M. Santos, C.~L. Benone, L.~C. Crispino, C.~A. Herdeiro, and E.~Radu,
  ``{Black holes with synchronised Proca hair: linear clouds and fundamental
  non-linear solutions},'' {\em JHEP}, vol.~07, p.~010, 2020.

\bibitem{PhysRevD.100.064032}
J.~Kunz, I.~Perapechka, and Y.~Shnir, ``Kerr black holes with parity-odd scalar
  hair,'' {\em Phys. Rev. D}, vol.~100, p.~064032, Sep 2019.

\bibitem{Kunz:2019sgn}
J.~Kunz, I.~Perapechka, and Y.~Shnir, ``{Kerr black holes with synchronised
  scalar hair and boson stars in the Einstein-Friedberg-Lee-Sirlin model},''
  {\em JHEP}, vol.~07, p.~109, 2019.

\bibitem{Collodel:2020gyp}
L.~G. Collodel, D.~D. Doneva, and S.~S. Yazadjiev, ``{Rotating
  tensor-multi-scalar-$N=2$ black holes},'' 7 2020.

\bibitem{Guo:2008hf}
Z.-K. Guo, N.~Ohta, and T.~Torii, ``{Black Holes in the Dilatonic
  Einstein-Gauss-Bonnet Theory in Various Dimensions. I. Asymptotically Flat
  Black Holes},'' {\em Prog. Theor. Phys.}, vol.~120, pp.~581--607, 2008.

\bibitem{Lee:2018zym}
B.-H. Lee, W.~Lee, and D.~Ro, ``{Expanded evasion of the black hole no-hair
  theorem in dilatonic Einstein-Gauss-Bonnet theory},'' {\em Phys. Rev. D},
  vol.~99, no.~2, p.~024002, 2019.

\bibitem{Gonzalez:2018aky}
G.~A. Gonzalez, B.~Kleihaus, J.~Kunz, and S.~Mojica, ``{Innermost stable
  circular orbits of neutron stars in dilatonic-Einstein-Gauss-Bonnet
  theory},'' {\em Phys. Rev. D}, vol.~99, no.~2, p.~024041, 2019.

\bibitem{Cunha:2016wzk}
P.~V. Cunha, C.~A.~R. Herdeiro, B.~Kleihaus, J.~Kunz, and E.~Radu, ``{Shadows
  of Einstein--dilaton--Gauss--Bonnet black holes},'' {\em Phys. Lett. B},
  vol.~768, pp.~373--379, 2017.

\bibitem{Blazquez-Salcedo:2017txk}
J.~L. Blazquez-Salcedo, F.~S. Khoo, and J.~Kunz, ``{Quasinormal modes of
  Einstein-Gauss-Bonnet-dilaton black holes},'' {\em Phys. Rev. D}, vol.~96,
  no.~6, p.~064008, 2017.

\bibitem{PhysRevLett.120.131103}
D.~D. Doneva and S.~S. Yazadjiev, ``New gauss-bonnet black holes with
  curvature-induced scalarization in extended scalar-tensor theories,'' {\em
  Phys. Rev. Lett.}, vol.~120, p.~131103, Mar 2018.

\bibitem{PhysRevLett.120.131104}
H.~O. Silva, J.~Sakstein, L.~Gualtieri, T.~P. Sotiriou, and E.~Berti,
  ``Spontaneous scalarization of black holes and compact stars from a
  gauss-bonnet coupling,'' {\em Phys. Rev. Lett.}, vol.~120, p.~131104, Mar
  2018.

\bibitem{PhysRevLett.120.131102}
G.~Antoniou, A.~Bakopoulos, and P.~Kanti, ``Evasion of no-hair theorems and
  novel black-hole solutions in gauss-bonnet theories,'' {\em Phys. Rev.
  Lett.}, vol.~120, p.~131102, Mar 2018.

\bibitem{PhysRevD.97.084037}
G.~Antoniou, A.~Bakopoulos, and P.~Kanti, ``Black-hole solutions with scalar
  hair in einstein-scalar-gauss-bonnet theories,'' {\em Phys. Rev. D}, vol.~97,
  p.~084037, Apr 2018.

\bibitem{PhysRevD.99.044017}
M.~Minamitsuji and T.~Ikeda, ``Scalarized black holes in the presence of the
  coupling to gauss-bonnet gravity,'' {\em Phys. Rev. D}, vol.~99, p.~044017,
  Feb 2019.

\bibitem{PhysRevD.99.064011}
H.~O. Silva, C.~F.~B. Macedo, T.~P. Sotiriou, L.~Gualtieri, J.~Sakstein, and
  E.~Berti, ``Stability of scalarized black hole solutions in
  scalar-gauss-bonnet gravity,'' {\em Phys. Rev. D}, vol.~99, p.~064011, Mar
  2019.

\bibitem{PhysRevD.99.064003}
A.~Bakopoulos, G.~Antoniou, and P.~Kanti, ``Novel black-hole solutions in
  einstein-scalar-gauss-bonnet theories with a cosmological constant,'' {\em
  Phys. Rev. D}, vol.~99, p.~064003, Mar 2019.

\bibitem{PhysRevD.99.104045}
D.~D. Doneva, K.~V. Staykov, and S.~S. Yazadjiev, ``Gauss-bonnet black holes
  with a massive scalar field,'' {\em Phys. Rev. D}, vol.~99, p.~104045, May
  2019.

\bibitem{PhysRevD.99.104041}
C.~F.~B. Macedo, J.~Sakstein, E.~Berti, L.~Gualtieri, H.~O. Silva, and T.~P.
  Sotiriou, ``Self-interactions and spontaneous black hole scalarization,''
  {\em Phys. Rev. D}, vol.~99, p.~104041, May 2019.

\bibitem{PhysRevD.98.104056}
D.~D. Doneva, S.~Kiorpelidi, P.~G. Nedkova, E.~Papantonopoulos, and S.~S.
  Yazadjiev, ``Charged gauss-bonnet black holes with curvature induced
  scalarization in the extended scalar-tensor theories,'' {\em Phys. Rev. D},
  vol.~98, p.~104056, Nov 2018.

\bibitem{PhysRevD.101.024033}
G.~Antoniou, A.~Bakopoulos, P.~Kanti, B.~Kleihaus, and J.~Kunz, ``Novel
  einstein--scalar-gauss-bonnet wormholes without exotic matter,'' {\em Phys.
  Rev. D}, vol.~101, p.~024033, Jan 2020.

\bibitem{PhysRevD.98.084011}
J.~L. Blazquez-Salcedo, D.~D. Doneva, J.~Kunz, and S.~S. Yazadjiev, ``Radial
  perturbations of the scalarized einstein-gauss-bonnet black holes,'' {\em
  Phys. Rev. D}, vol.~98, p.~084011, Oct 2018.

\bibitem{Blazquez-Salcedo:2020rhf}
J.~L. Blazquez-Salcedo, D.~D. Doneva, S.~Kahlen, J.~Kunz, P.~Nedkova, and S.~S.
  Yazadjiev, ``{Axial perturbations of the scalarized Einstein-Gauss-Bonnet
  black holes},'' {\em Phys. Rev. D}, vol.~101, no.~10, p.~104006, 2020.

\bibitem{Blazquez-Salcedo:2020caw}
J.~L. Blazquez-Salcedo, D.~D. Doneva, S.~Kahlen, J.~Kunz, P.~Nedkova, and S.~S.
  Yazadjiev, ``{Polar quasinormal modes of the scalarized Einstein-Gauss-Bonnet
  black holes},'' {\em Phys. Rev. D}, vol.~102, no.~2, p.~024086, 2020.

\bibitem{Doneva:2020qww}
D.~D. Doneva, K.~V. Staykov, S.~S. Yazadjiev, and R.~Z. Zheleva,
  ``{Multi-scalar Gauss-Bonnet gravity -- hairy black holes and
  scalarization},'' 6 2020.

\bibitem{Doneva:2017duq}
D.~D. Doneva and S.~S. Yazadjiev, ``{Neutron star solutions with curvature
  induced scalarization in the extended Gauss-Bonnet scalar-tensor theories},''
  {\em JCAP}, vol.~04, p.~011, 2018.

\bibitem{KLEIHAUS2020135401}
B.~Kleihaus, J.~Kunz, and P.~Kanti, ``Particle-like ultracompact objects in
  einstein-scalar-gauss-bonnet theories,'' {\em Physics Letters B}, vol.~804,
  p.~135401, 2020.

\bibitem{PhysRevD.102.024070}
B.~Kleihaus, J.~Kunz, and P.~Kanti, ``Properties of ultracompact particlelike
  solutions in einstein-scalar-gauss-bonnet theories,'' {\em Phys. Rev. D},
  vol.~102, p.~024070, Jul 2020.

\bibitem{PhysRevLett.123.011101}
P.~V.~P. Cunha, C.~A.~R. Herdeiro, and E.~Radu, ``Spontaneously scalarized kerr
  black holes in extended scalar-tensor--gauss-bonnet gravity,'' {\em Phys.
  Rev. Lett.}, vol.~123, p.~011101, Jul 2019.

\bibitem{Collodel_2020}
L.~G. Collodel, B.~Kleihaus, J.~Kunz, and E.~Berti, ``Spinning and excited
  black holes in einstein-scalar-gauss{\textendash}bonnet theory,'' {\em
  Classical and Quantum Gravity}, vol.~37, p.~075018, mar 2020.

\bibitem{Dima:2020yac}
A.~Dima, E.~Barausse, N.~Franchini, and T.~P. Sotiriou, ``{Spin-induced black
  hole spontaneous scalarization},'' 6 2020.

\bibitem{Hod:2020jjy}
S.~Hod, ``{Onset of spontaneous scalarization in spinning Gauss-Bonnet black
  holes},'' 6 2020.

\bibitem{Dima:2020rzg}
A.~Dima and E.~Barausse, ``{Numerical investigation of plasma-driven
  superradiant instabilities},'' {\em Class. Quant. Grav.}, vol.~37, p.~175006,
  2020.

\bibitem{Racz:2011qu}
I.~Racz and G.~Z. Toth, ``{Numerical investigation of the late-time Kerr
  tails},'' {\em Class. Quant. Grav.}, vol.~28, p.~195003, 2011.

\bibitem{Gao:2018acg}
Y.-X. Gao, Y.~Huang, and D.-J. Liu, ``{Scalar perturbations on the background
  of Kerr black holes in the quadratic dynamical Chern-Simons gravity},'' {\em
  Phys. Rev. D}, vol.~99, no.~4, p.~044020, 2019.

\bibitem{Krivan:1996da}
W.~Krivan, P.~Laguna, and P.~Papadopoulos, ``{Dynamics of scalar fields in the
  background of rotating black holes},'' {\em Phys. Rev. D}, vol.~54,
  pp.~4728--4734, 1996.

\bibitem{Krivan:1997hc}
W.~Krivan, P.~Laguna, P.~Papadopoulos, and N.~Andersson, ``{Dynamics of
  perturbations of rotating black holes},'' {\em Phys. Rev. D}, vol.~56,
  pp.~3395--3404, 1997.

\bibitem{Zenginoglu:2010cq}
A.~Zenginoglu, ``{Hyperboloidal layers for hyperbolic equations on unbounded
  domains},'' {\em J. Comput. Phys.}, vol.~230, pp.~2286--2302, 2011.

\bibitem{Harms:2013ib}
E.~Harms, S.~Bernuzzi, and B.~Brügmann, ``{Numerical solution of the 2+1
  Teukolsky equation on a hyperboloidal and horizon penetrating foliation of
  Kerr and application to late-time decays},'' {\em Class. Quant. Grav.},
  vol.~30, p.~115013, 2013.

\bibitem{Harms:2014dqa}
E.~Harms, S.~Bernuzzi, A.~Nagar, and A.~Zenginoglu, ``{A new gravitational wave
  generation algorithm for particle perturbations of the Kerr spacetime},''
  {\em Class. Quant. Grav.}, vol.~31, no.~24, p.~245004, 2014.

\bibitem{Kruger:2019zuz}
C.~Krüger and K.~Kokkotas, ``{Fast Rotating Relativistic Stars: Spectra and
  Stability without Approximation},'' 10 2019.

\bibitem{Kruger:2020ykw}
C.~Krüger and K.~Kokkotas, ``{Dynamics of Fast Rotating Neutron Stars: An
  Approach in the Hilbert Gauge},'' 8 2020.

\bibitem{RuoffPhD}
J.Ruoff, {\em The Numerical Evolution of Neutron Star Oscillations}.
\newblock PhD thesis, University ot Tuebingen, 2000.

\bibitem{Berti:2009kk}
E.~Berti, V.~Cardoso, and A.~O. Starinets, ``{Quasinormal modes of black holes
  and black branes},'' {\em Class. Quant. Grav.}, vol.~26, p.~163001, 2009.

\bibitem{Berti:2005ys}
E.~Berti, V.~Cardoso, and C.~M. Will, ``{On gravitational-wave spectroscopy of
  massive black holes with the space interferometer LISA},'' {\em Phys. Rev.
  D}, vol.~73, p.~064030, 2006.

\end{thebibliography}

\end{document}